\documentstyle[12pt,dina4p,epsfig]{article}
\newcommand{\be}{\begin{eqnarray}}
\newcommand{\ee}{\end{eqnarray}}
\newcommand{\nn}{\nonumber}

\renewcommand{\d}{\mbox{d}}
\newcommand{\dis}{\displaystyle}
\newcommand{\g}{\gamma}

\title{
  \vskip-2cm
  {\baselineskip16pt
    \centerline{\normalsize \tt DESY 96-223  \hfill ISSN 0418-9833}
    \centerline{\normalsize \tt hep-ph/9610489 \hfill}
    \centerline{\normalsize \tt October 1996 \hfill}
  }
  \vskip2cm
  {\bf 
    Inclusive Two--Jet Production in Photon--Photon Collisions:\\
    Direct and Resolved Contributions in Next--to--Leading Order QCD
  }
  \author{
    {T. Kleinwort\footnote{E--mail:\ tkleinw@ifh.de, present
     address:\ DESY--IfH Zeuthen, Platanenallee 6, 15735 Zeuthen},
     G. Kramer} \\ 
    {II. Institut f\"ur Theoretische Physik}\thanks
     {Supported by Bundesministerium f\"ur Forschung und
     Technologie, Bonn, Germany under Contract 05\,7HH92P(5) and
     EEC Program ``Human Capital and Mobility'' through Network
     ``Physics at High Energy Colliders'' under Contract
     CHRX--CT93--0357 (DG12 COMA) }
    \\
    {Universit\"at Hamburg} \\
    {D - 22761 Hamburg, Germany}
  }
  \date{}
}
\begin{document}
\maketitle
\vspace{3cm}
\begin{abstract}
\thispagestyle{empty}
We have calculated inclusive two--jet production in photon--photon
collisions superimposing direct, single--resolved and double--resolved
cross sections for center--of--mass energies of TRISTAN and LEP1.5. All
three contributions are calculated up to next--to--leading order. The
results are compared with recent experimental data. Three NLO sets of
parton distributions of the photon are tested.
\end{abstract}
\newpage
\setcounter{page}{1}
\section{Introduction}
The production of high-$p_T$ jets probes the short-distance dynamics
of photon-photon reactions. In addition to providing tests of
perturbative QCD, $\g\g$ processes with real or almost real photons
give us information on the photon structure function which is
complementary to the information gained from deep inelastic scattering
on a real photon. The latter process essentially probes the quark
distribution while high-$p_T$ jet production is also sensitive to the
gluon distribution of the photon.

As it is well known high-$p_T$ processes induced by real photons have
a rather complex structure. This arises from the fact that the photon
couples to the hard subprocess either directly or through its quark
and gluon content. 

In leading order QCD (LO) the cross section for the production of
jets can therefore be decomposed into a direct, single-resolved,
and double-resolved component. In next-to-leading order (NLO) the
photon-quark collinear singularities in the direct and single
resolved components are subtracted and absorbed into the quark
distribution of the photon in accord with the factorization theorem.
This subtraction at the factorization scale $M$ introduces an
interdependence of the three components so that the separation
into the direct, single-resolved, and double-resolved contributions
becomes scale dependent. This means that in NLO all three components
must be considered together and consistently be calculated up to
NLO in the photon structure functions and in the hard scattering
cross sections.

Complete NLO calculations have been done previously for the case of the
inclusive single--jet cross section \cite{ref1, ref2, ref3} and
compared to experimental data from TRISTAN \cite{ref4, ref5}. Satisfactory
agreement between data and theory has been found. In our previous work
\cite{ref2, ref3} we also calculated the NLO inclusive two-jet cross
sections for the direct and single-resolved contribution and
estimated the double-resolved contribution by the LO calculation
with $k$ factors taken from the NLO inclusive one-jet cross section.
In the meantime the calculation of the NLO corrections for the
double-resolved cross section has been completed also for the
inclusive two-jet cross section using techniques also employed
for the direct and single-resolved cross sections in our previous
work \cite{ref2, ref3}.

In this paper we shall present the complete NLO predictions for the
inclusive two-jet cross section incorporating all three components,
direct, single-resolved and double-resolved. We shall compare
these predictions with the older TRISTAN data \cite{ref4, ref5} and with
recent data from LEP at $e^+e^-$ energies of $130$ and $136\,GeV$
(LEP1.5) \cite{ref6}. The details of the NLO calculation of the
double-resolved cross section are given elsewhere \cite{ref7}. Those of the
direct and single-resolved cross section are described in \cite{ref3}.

The outline of the rest of the paper is as follows. In section 2
we explain the formalism used to calculate the inclusive two-jet
cross section. Section 3 contains our results for the cross sections
and some studies on the overall scale dependence and the dependence
of the two-jet cross section on the cone radius R. Here we also
present the results for different sets of photon structure
functions and the comparison with the TRISTAN and LEP1.5 data. Section
4 contains a short summary.

\section{Inclusive Two-Jet Cross Sections}

As explained in the introduction inclusive jet cross sections at
large $p_T$ receive contributions from three parts according to how the
incoming photons take part in the hard scattering subprocesses.
For example, when both photons are resolved the inclusive two-jet
cross section has the following form
\be
 \frac{\d^3\sigma^{DR}}{\d p_T\d\eta_1\d\eta_2}&=&
 \sum\limits_{i,j=q,g}\int\limits\!\d x_1\int\limits\!\d x_2\,
 F_{i/\g}(x_1,M)\,F_{j/\g}(x_2,M)\nn\\
 &&\left(\left(\frac{\d^3\sigma^{DR}(ij\to\mbox{jet}_1+\mbox{jet}_2)}
              {\d p_T\d\eta_1\d\eta_2}\right)_{\mbox{LO}}
 +\frac{\alpha_s(\mu)}{2\pi}K_{ij}^{DR}(R,M,\mu)\right)
 \label{gl1}
\ee
In (\ref{gl1}), $p_T$ is the transverse momentum of the measured or trigger
jet with rapidity $\eta_1$. $\eta_2$ denotes the rapidity of another
jet such that in the three-jet sample (in the NLO calculation only up
to three jets occur in the final state, the photon remnant jets are
not counted) these two jets have the largest and second largest $p_T$
$({p_T}_1,{p_T}_2>{p_T}_3)$. 
The first term on the right-hand side of (\ref{gl1}) stands for the LO
cross section and the second term is the NLO correction which
depends on the factorization scale $M$ at which the initial state
singularities are absorbed into the parton distributions of the
photon. $\mu$ is the renormalization scale. The variable $R$ is the
usual cone size parameter which defines the size of the jets with
transverse momentum $p_T$, rapidity $\eta$ and azimuthal angle $\phi$.
When two partons fulfill the Snowmass constraint \cite{ref8} with
the cone size parameter $R$ they are recombined in one jet. The same
jet definition is used in the analysis of the experimental data
by the TOPAZ \cite{ref4}, AMY \cite{ref5} and OPAL \cite{ref6}
collaborations. It is clear that either of the two jets can consist
of a recombination of two partons inside a cone with radius $R$.

The NLO corrections in the hard scattering processes are calculated
with the phase space slicing method using for the separation of the
$\g\g\to2$ jets and $\g\g\to3$ jets cross sections an invariant mass
cut-off $y$, defined as $2p_ip_j<ys$, where $s$ is the partonic
center-of-mass energy squared. For example, the cross section for
the direct process $\g\g\to q\bar{q}g$ has soft, initial, and final
state collinear singularities. The $\g\g\to q\bar{q}g$ cross section is
integrated over these singular regions up to the cut-off $y$ which
isolates the respective singularities. These cancel against the
singular contributions of the virtual corrections to $\g\g\to q\bar{q}$
and the subtraction terms at the scale $M$ to be absorbed into
the parton distribution functions of the photon. Outside the
cut-off region controlled by the parameter $y$ we have genuine
$q\bar{q}g$ final states. In this contribution two of the partons are
combined if they obey the Snowmass constraint. The LO contribution,
the NLO virtual contributions, and
the NLO corrections inside the $y$ cut contribute to the two-jet cross 
section together with the contributions inside the cone with
radius $R$. The part of the $q\bar{q}g$ cross section  not fulfilling
the cone recombination condition is the 3-jet cross section from which
we have calculated the inclusive 2-jet cross section as a function of 
$p_T$, $\eta_1$ and $\eta_2$. 

For the exclusive 2-jet cross section ${p_T}_1={p_T}_2=p_T$.
To separate double singularities in the $q\bar{q}g$ contributions we
have used the method of partial fractioning. The same partial
fractioned expressions are applied for the calculation inside and
outside the $y$ cut. Inside the $y$ cut the calculation is done
analytically in the approximation that terms $O(y)$ are
neglected. Outside the $y$ cut the contributions to the inclusive cross
section are evaluated numerically with no further 
approximations. Because of the approximation in the analytic part,
the parameter $y$ must be chosen very small, of the order of
$10^{-3}$ to $10^{-4}$. For such values of $y$ the inclusive cross
section should be independent of the invariant mass parameter
$y$. However the inclusive two-jet cross section depends on the cone
size $R$, which must be chosen as in the analysis of the experimental
data. In the case of the single-- and double--resolved contributions
the calculation of the NLO corrections proceeds in the same way. For
example, in the single--resolved case, we have in LO the contributions
from $\g g\to q\bar{q}$ and $\g q\to qg$ and in NLO we must consider
$\g g\to q\bar{q}g, \g q\to qq\bar{q}$ and $\g q\to qgg$ which are
treated in the same fashion as the 3--parton contributions 
in the direct case. In the calculation of the double--resolved cross
section we have no photons in the initial state, only quarks and
gluons, and we have many more channels to consider as processes
with three partons in the final state, which will not be given
here. Otherwise the calculation proceeds in a completely analogous way
to the direct and single--resolved case. 

Before the final results were
obtained some tests of the NLO corrections have been performed. First
we checked that the cross sections are independent of the cut-off $y$
if $y$ is chosen small enough. This was the case for $y\le10^{-3}$
in all considered cases. For $y>10^{-3}$ we observed some small $y$
dependence which is caused by the approximation in the analytical
part. Second the inclusive one--jet cross section for the
double--resolved  contribution was calculated and compared with our
earlier results for which a completely different method for the
cancellation of the infrared and collinear singularities was applied
\cite{ref2, ref3}. Very good agreement was found in all
channels. Third we tested that the sum of the NLO direct and 
the LO single--resolved cross section is independent of the
factorization scale $M$. The same test was performed for the sum of the
NLO single--resolved cross section and the LO double--resolved cross
section. Details of these tests can be found in \cite{ref7}. In this
work also the details of the analytic calculation and the
cancellation of the singularities for the NLO corrections of the
double--resolved cross section are given. For the direct and the
single--resolved cross sections the calculation of the NLO
corrections was described earlier in \cite{ref3} and \cite{ref9},
respectively, and also in \cite{ref7}.

For the calculation of the single-- and double--resolved cross sections
we need the parton distributions of the photon $F_{i/\g}$
which appear in (\ref{gl1}) and in the equivalent formula for the
single-resolved cross section (see \cite{ref2}). We have chosen the
NLO set of Gl\"uck, Reya and Vogt (GRV) in the
$\overline{\mbox{MS}}$ scheme 
\cite{ref10} as our standard set. For a comparison we shall 
employ also the $\overline{\mbox{MS}}$ sets of Gordon
and Storrow \cite{ref11} and of 
Aurenche et al. \cite{ref12}. We choose all scales $\mu=M=p_T$
except when we test the scale dependence. $\alpha_s$ is calculated
from the two--loop formula with $N_f=4\,(5)$ massless flavours with
$\Lambda_{\overline{\mbox{\tiny{MS}}}}^{(4)}=0.20\,GeV
(\Lambda_{\overline{\mbox{\tiny{MS}}}}^{(5)}=0.13\,GeV)$ 
for the $e^+e^-$ center of mass energy $\sqrt{S}=58\,GeV
(\sqrt{S}=133\,GeV)$. The charm, respectively, bottom quarks are also
treated as light flavours with the boundary condition that the charm
(bottom) content of the photon structure function vanishes for
$M^2<m_c^2\,(m_b^2)$ with $m_c=1.5\,GeV (m_b=5.0\,GeV)$ for the GRV
parton distributions \cite{ref10}. For the other two sets the 
treatment of heavy flavours is somewhat different. The details are
found in the respective references \cite{ref11, ref12}.

\section{Results and Comparison with Data from TOPAZ, AMY, and OPAL}

In $e^+e^-$ collisions the jets are produced via the exchange of two
almost real photons. The spectrum of these small $Q^2$ photons is
described by the Weizs\"acker-Williams approximation \cite{ref13}
\be
   F_{\g/e}(z,E)&=&\frac{\alpha}{2\pi}\frac{1+(1-z)^2}{z}
   \left\{\ln\left(\frac{E^2\theta_{\max}^2(1-z)^2+m_e^2z}{m_e^2z^2}\right)
   \right.\nn\\
   &&+\left.2(1-z)\left(\frac{m_e^2z}{E^2\theta_{\max}^2(1-z)^2+m_e^2z^2}
   -\frac12\right)\right\}\label{gl2}
\ee
where $\theta_{\max}$ is the maximally allowed angle of the electron
(positron) which is given experimentally by the anti-tagging
counters. $E$ is the beam energy of the electron (positron) and
$z=E_\gamma/E$ is the fraction of the electron (positron) energy
transferred to the respective photon.

Then the inclusive two-jet cross section for the reaction
$e^++e^-\to e^++e^-+\mbox{jet}_1+\mbox{jet}_2+X$ is calculated from 
\be
 \frac{\d^3\sigma(e^++e^-\to e^++e^-+\mbox{jet}_1+\mbox{jet}_2+X)}
      {\d p_T\d\eta_1\d\eta_2}&=&\nn\\
 &&\!\!\!\!\!\!\!\!\!\!\!\!\!\!\!\!\!\!\!\!\!\!\!\!\!\!\!\!
   \!\!\!\!\!\!\!\!\!\!\!\!\!\!\!\!\!\!\!\!\!\!\!\!\!\!\!\!
   \!\!\!\!\!\!\!\!\!\!\!\!\!\!\!\!\!\!\!\!\!\!\!\!\!\!\!\!
   \int\limits\!\d z_1\int\limits\!\d z_2\,
   F_{\g/e}(z_1,E)\,F_{\g/e}(z_2,E)
   \frac{\d^3\sigma(\g+\g\to\mbox{jet}_1+\mbox{jet}_2+X)}
      {\d p_T\d\eta_1\d\eta_2}\label{gl3}
\ee
where the cross section on the right-hand side is for the
interaction of two real photons $(Q^2=0)$ and is obtained from
(\ref{gl1}) for the double-resolved case and similar formulas for the
direct and single-resolved $\g\g$ cross section (see \cite{ref2}).

First we show in Fig.\ 1 the dependence of the inclusive 2--jet
cross section on the cone radius $R$ for the special case of LEP2
kinematics, $\sqrt{S}=175\,GeV, \theta_{\max}=1.72^\circ$,  
since we cannot expect this dependence to be tested with
TRISTAN or LEP1.5 data. Of course, the LO cross section is
independent of $R$ which is shown as the broken line in Fig.\ 1. In
the NLO calculation the cross section increases with increasing $R$,
first almost linearly in $\ln R$ for $R$ values up to 0.75 and somewhat
stronger for $R>0.75$. At $R\simeq0.9$ the LO and NLO cross sections
are equal for the special kinematical point $p_T=10\,GeV, \eta_1=\eta_2=0$.
Therefore for $R$ values, chosen between 0.7 and 1.0, the higher order
corrections are moderate. The LO curve is obtained with the two-loop
formula with the same $\Lambda_{\overline{\mbox{\tiny{MS}}}}$
value and with the same NLO 
photon structure functions from GRV. In a genuine LO calculation with
one--loop $\alpha_s$ and LO structure functions the cross section is
somewhat larger. That the NLO corrections near $R = 1$ are not very
large is due to 
the dominance of the direct contribution which is not changed by
the NLO corrections very much \cite{ref2, ref3} in contrast to the NLO
corrections for the double--resolved cross section \cite{ref7}.

In Fig.\ 2, 3, and 4 the inclusive two-jet cross sections
$\frac{\dis\d^3\sigma}{\dis\d p_T\d\eta_1\d\eta_2}$ are plotted as a functions
of $p_T$, integrated over the rapidities $\eta_1$ and $\eta_2$
as in the experimental analysis of the TOPAZ \cite{ref4}, AMY \cite{ref5}
and OPAL \cite{ref6} collaborations.  The conditions for the TOPAZ
experiment are $\theta_{\max}=3.2^{\circ}, z_1, z_2\le0.75$ and
$|\eta_1|,|\eta_2|\le0.7$. The cone radius $R=1$ is used in all
three figures. The comparison with the TOPAZ data \cite{ref4} is seen in
Fig.\ 2. The full curve is the sum of the direct, single--resolved and
double--resolved contributions. The separate cross sections are
also plotted. From these curves one can see that at the TRISTAN
energy the resolved part is significant only at very small $p_T$.
At large $p_T\simeq8\,GeV$ the direct contribution is dominant. But
both, the double--resolved and even more the single--resolved cross
section make a non--negligible contribution. The agreement of the
theoretical prediction with the TOPAZ data is very good. It is clear
from this comparison that the data cannot be explained by the direct
contribution alone even at the larger $p_T$. In Fig.\ 3 the same
curves for the AMY experiment are shown. In this experiment the cuts
are $\theta_{\max}=13^{\circ}$ and $|\eta_1|,|\eta_2|\le1$.
$z_1$ and $z_2$ are not restricted further. We show again the three
contributions, the 
direct, single--resolved and double--resolved cross section. The
data points must be compared with the full curve. The agreement
with the theoretical curve is again quite good. The AMY cross section
is larger than the TOPAZ cross section, essentially since
$\theta_{\max}$ is larger. In the fall of 1995 LEP was run at
center-of-mass energies of $130$ and $136\,GeV$. During this rather
short run the OPAL collaboration was able to measure inclusive jet
cross sections and analyze them with the cone jet finding
algorithm \cite{ref6}. The cone radius was also $R=1$. In Fig.\ 4 we
show their data of the inclusive two-jet cross section together with
our prediction (full curve) and the cross sections for the direct,
single--resolved and  double--resolved part. The kinematical
constraints are $\theta_{\max}=1.43^{\circ}, |\eta_1|,|\eta_2|\le1$.
Because of the larger c.m.\ energy we have taken $N_f=5$ and the
corresponding
$\Lambda_{\overline{\mbox{\tiny{MS}}}}=0.130\,GeV$. At small 
$p_T$ the data are very accurate. The combined statistical and systematic
errors are shown. The agreement of the theory with the data is
good over the whole $p_T$ range. We see that the double--resolved
contribution is much more important at this higher energy. It gives
the dominant contribution for $p_T\le5\,GeV$. Even at $p_T=16\,GeV$
the direct contribution is only 60\% of the total. Except for the last
high $p_T$ point which has a larger error the pure direct curve is not
consistent with the data. This shows that the other two contributions,
the single--resolved and double--resolved, are needed to explain the
data. The OPAL collaboration measured also
the inclusive single--jet cross section. The comparison of their data
with our predictions is shown in \cite{ref6} and \cite{ref7}. The
agreement between theory and the data is as good as in the two-jet
case. The inclusive two-jet cross section for LEP1.5 contains an extra
factor 2 as compared to the definition used for the TRISTAN energy in
order to account for the definition as used in \cite{ref6}.

In all the comparisons between experiment and theory shown above, we employed
the GRV presentation of the parton distributions of the photon.
Other NLO parton distributions are those of Gordon and Storrow
\cite{ref11} and Aurenche et al.\ \cite{ref12}. Before we compare the data
with predictions based on these other parton distributions we
investigate the scale dependence of the inclusive two-jet cross
sections for the two $e^+e^-$ c.m.\ energies corresponding to TRISTAN
and LEP1.5. The results are shown in Fig.\ 5a,b where we have plotted
the two-jet cross sections for three different scales $\mu=M=\xi p_T$
with $\xi=1/2$, $1$ and $2$, and compared them with the TOPAZ (Fig.\ 5a)
and OPAL data (Fig.\ 5b). In both cases the cross section is remarkably
independent of the scale change. Only at rather small $p_T<4\,GeV$
the cross section in Fig.\ 5a exhibits some scale change. For
$ p_T>4\,GeV$ the cross section at $\sqrt{S}=58\,GeV$ deviates less
than $\pm4\%$ from the standard with $\xi=1$.
Similarly the cross section at $\sqrt{S}=133\,GeV$ shown in Fig.\ 5b
varies by less than $\pm2\%$ above $p_T=8\,GeV$. This small scale change
of the $\g\g$ jet cross sections (the inclusive one--jet cross section shows a
similar behaviour) results from the fact that the scale change
originating from the change of the renormalization scale $\mu$ leading
to a decrease of the cross section with increasing $\xi$
is to a large extend cancelled by the dependence on the
factorization scale $M$, which is opposite to the change in $\mu$. The
dependence on $M$ is mostly cancelled between the NLO direct and the LO
single-resolved cross section and similarly between NLO
single-resolved and LO double-resolved cross section \cite{ref7}.
But some dependence on $M$ remains when all three cross sections are
evaluated in NLO. Thus this reduced scale dependence of the
$\g\g$ jet cross sections would be a good place to determine
$\Lambda_{\overline{\mbox{\tiny{MS}}}}$. For this we need
photon structure functions with varying
$\Lambda_{\overline{\mbox{\tiny{MS}}}}$ which are not available
yet and accurate  data in particular for large $p_T$. In order to see
the change of the two--jet cross section when we change from
$\Lambda_{\overline{\mbox{\tiny{MS}}}}^{(5)}=0.130\,GeV$ to
$\Lambda_{\overline{\mbox{\tiny{MS}}}}^{(5)}=0.250\,GeV$ in the
formula for $\alpha_s$ we have plotted the cross section for this
large value of $\Lambda_{\overline{\mbox{\tiny{MS}}}}$ also in
Fig.\ 5b. As to be expected the cross section is larger now. At the
smaller $p_T<5\,GeV$ the data points lie below the theoretical
curve. Since $\Lambda_{\overline{\mbox{\tiny{MS}}}}$ was not
changed in the photon structure function consistently we  do not draw any
conclusions from this observation. At large $p_T$ the data points are still
consistent with the theoretical curve for the larger
$\Lambda_{\overline{\mbox{\tiny{MS}}}}^{(5)}$. The equivalent
result at $\sqrt{S}=58\,GeV$  is shown in Fig.\ 5a, where we compare
the results for
$\Lambda_{\overline{\mbox{\tiny{MS}}}}^{(4)}=0.200\,GeV$ with
$\Lambda_{\overline{\mbox{\tiny{MS}}}}^{(4)}=0.358\,GeV$ which
is equivalent to
$\Lambda_{\overline{\mbox{\tiny{MS}}}}^{(5)}=0.250\,GeV$ when
we choose four active flavours. Due to the larger experimental errors the
curve with the larger $\Lambda_{\overline{\mbox{\tiny{MS}}}}$
value is still compatible with the 
data. At the larger $p_T$ the agreement is even somewhat
improved. This shows that with the limited experimental accuracy of
the data at both energies it seems difficult to draw clear
conclusions even if the photon structure functions were given for
varying $\Lambda_{\overline{\mbox{\tiny{MS}}}}$.

In Fig.\ 6a,b we investigated the influence of the gluon distribution
of the photon on the two--jet cross section. It is clear that the gluon
distribution will have a remarkable influence only at small
$p_T$ since this distribution function is dominant for small $x$. To
see this we have assumed $F_{g/\g}=0$ with the assumption that the quark
distribution functions are already well enough constrained by the deep
inelastic $e$--$\g$ scattering data. As is seen in Fig.\ 6a,b the
prediction with $F_{g/\g}=0$ gives very bad agreement with the data at
small $p_T$ ($p_T<5\,GeV$ in Fig.\ 6a and $p_T<8\,GeV$ in
Fig.\ 6b). From this we conclude that the two-jet data require a 
non-vanishing $F_{g/\g}$ in the photon structure function. This agrees
with earlier findings in \cite{ref4, ref5} on the basis of more model
dependent calculations and with LO calculations in \cite{ref14} based on
new GS structure functions.

The last point is the prediction for other parton distributions of
the photon than the GRV distribution in NLO. The comparison is
shown in Fig.\ 7a,b together with the TOPAZ and OPAL experimental
data. As we can see there is little difference between the results
for the GRV \cite{ref10} and the ACFGP \cite{ref12} structure functions.
The cross sections for the GS \cite{ref11} distribution functions are
somewhat smaller than for GRV. In average, for both c.m. energies,
the data agree better with the GRV and ACFGP curve than with the curve
based on the GS structure function, in particular in the comparison
with the TOPAZ data. In order to exclude one of these structure
functions much more accurate data are needed. In general, we
conclude, that all three NLO parton distributions of the photon
account remarkably well for the existing data of the inclusive
two-jet production. It is clear, that additional information,
as for example, the rapidity distributions for the two jets could
constrain these structure functions further.

The same comparison has been done with the AMY two-jet data. Since
in this case, the GRV curve lies at small $p_T$ above the data
points the agreement with the prediction with the GS structure
function is better than for the TOPAZ data. 

Overall we can state that the agreement between the theoretical
predictions and the measurements of the two--jet cross sections is
remarkably good. This is true for the magnitude of the cross section
and its shape as a function of $p_T$. We should not conceal, however,
that the NLO calculation gives the jet cross sections for massless
partons, whereas the experimental jet cross sections are measured from
hadrons. Assuming the usual parton--hadron duality, no difference
between parton and hadron level would be observed. This parton--hadron
duality is very well obeyed in jet production for $e^+e^-$
annihilation at the $Z$ resonance. In this case the hard scale is
larger than in the $\g\g$ processes, so that it is problematic to draw
any firm conclusions from it. On the other hand it is not easy to
estimate the hadronization effects for jet production in $\g\g$
processes. For this one needs a Monte Carlo routine which generates
the final state partons including the NLO correction with subsequent
fragmentation into hadron jets. Such a routine is not available
yet. Estimates on the basis of LO Monte Carlos may not be too
conclusive. Further details for this case are reported in \cite{ref6}.

\section{Summary}

Differential inclusive two--jet cross sections $\d\sigma/\d p_T$ have been
calculated as a function of $p_T$ in NLO for the direct,
single--resolved and double--resolved contributions. The superposition
of these cross sections are compared to the TOPAZ, AMY and OPAL
data. We obtained good agreement between measured data and the
theoretical predictions. The same is true for the single inclusive
cross sections \cite{ref6,ref7}. The single-- and double--resolved
cross sections are obtained with the GRV \cite{ref10} photon structure
function. Two other sets GS \cite{ref11} and ACFGP \cite{ref12} were
also investigated. With them the data are described almost equally
well. The $\g\g$ jet cross sections are remarkably stable against
scale changes. With increasing center--of--mass energy the
double--resolved contribution becomes more and more important. This
is clearly seen by comparing the prediction for the TRISTAN and LEP1.5
energies. Similar to earlier findings we exclude a description of the
data without a gluon distribution in the photon. It is hoped that more
accurate data from LEP2 may be able to constrain the parton
distributions of the photon further.

\section{Acknowledgements}

We thank S.~S\"oldner--Rembold for information about the OPAL
measurements and for communicating the data before publication.
We are grateful to M.~Klasen for reading the manuscript.

\newpage
\section{Figure Captions}
\begin{enumerate}
 
\item The inclusive two--jet cross section as a function of the cone
      radius $R$ for LEP2 kinematics and $p_T=10\,GeV,\eta_1=\eta_2=0$.

\item Inclusive two--jet cross section $\d\sigma/\d p_T$ with $R=1$ as
      a function of $p_T$ for direct (dashed), single--resolved
      (dashed--dotted) and double--resolved (dotted) production in
      NLO. The full curve is the sum of all three components compared to
      TOPAZ data $(|\eta_1|,|\eta_2|<0.7)$

\item Inclusive two--jet cross section $\d\sigma/\d p_T$ with $R=1$ as
      a function of $p_T$ for direct (dashed), single--resolved
      (dashed--dotted) and double--resolved (dotted) production in
      NLO. The full curve is the sum of all three components compared to
      AMY data $(|\eta_1|,|\eta_2|<1)$

\item Inclusive two--jet cross section $\d\sigma/\d p_T$ with $R=1$ as
      a function of $p_T$ for direct (dashed), single--resolved
      (dashed--dotted) and double--resolved (dotted) production in
      NLO. The full curve is the sum of all three components compared to
      OPAL data $(|\eta_1|,|\eta_2|<1)$. 

\item Inclusive two--jet cross sections $\d\sigma/\d p_T$ for three
      different scales $\mu=M=\xi p_T$ with $\xi=\frac12$ (dashed),
      $\xi=1$ (full), $\xi=2$ (dotted) and different
      $\Lambda_{\overline{\mbox{\tiny{MS}}}}$ (dashed--dotted)
      compared to (a) TOPAZ data
      $(\Lambda_{\overline{\mbox{\tiny{MS}}}}^{(4)}=0.358\,GeV)$ and 
      (b) OPAL data
      $(\Lambda_{\overline{\mbox{\tiny{MS}}}}^{(5)}=0.250\,GeV)$.  

\item Inclusive two--jet cross section $\d\sigma/\d p_T$ with GRV
      parton distributions and with $F_{g/\g}=0$ (dashed) and
      $F_{g/\g}\ne0$ (full) compared to (a) TOPAZ data and (b) OPAL data.

\item Inclusive two--jet cross section $\d\sigma/\d p_T$ for different
        parton distribution sets: GRV \cite{ref10} (full), GS
        \cite{ref11} (dashed) and ACFGP \cite{ref12} (dotted) 
        compared to (a) TOPAZ and (b) OPAL two-jet data.

\end{enumerate}
\begin{figure}[ht]
 \begin{picture}(17,10)
  \put(0,0)
   {\epsfig{file=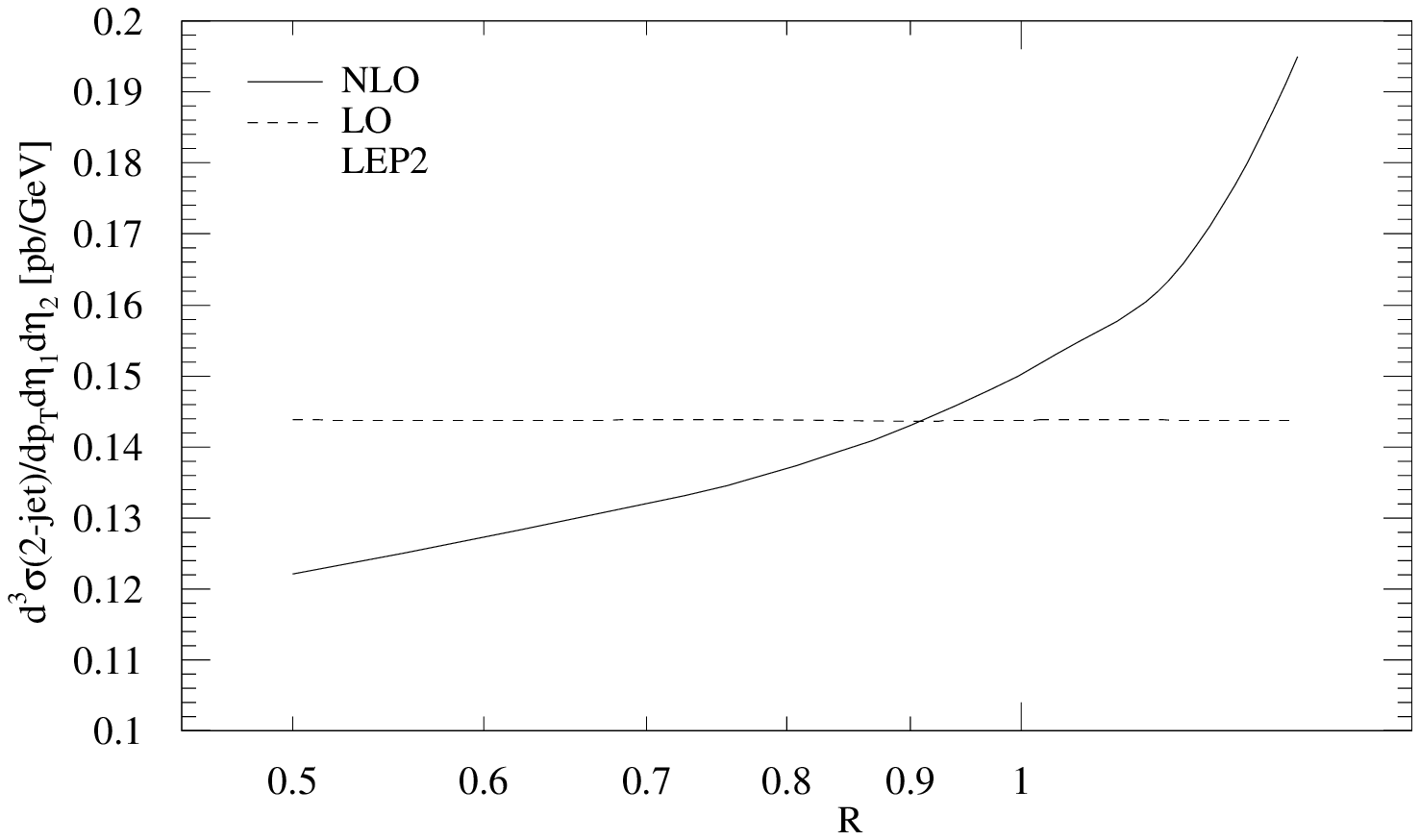, height=10cm}}
 \end{picture}
 \caption{\ }
\end{figure}
\begin{figure}[bh]
 \begin{picture}(17,10)
  \put(0,0)
   {\epsfig{file=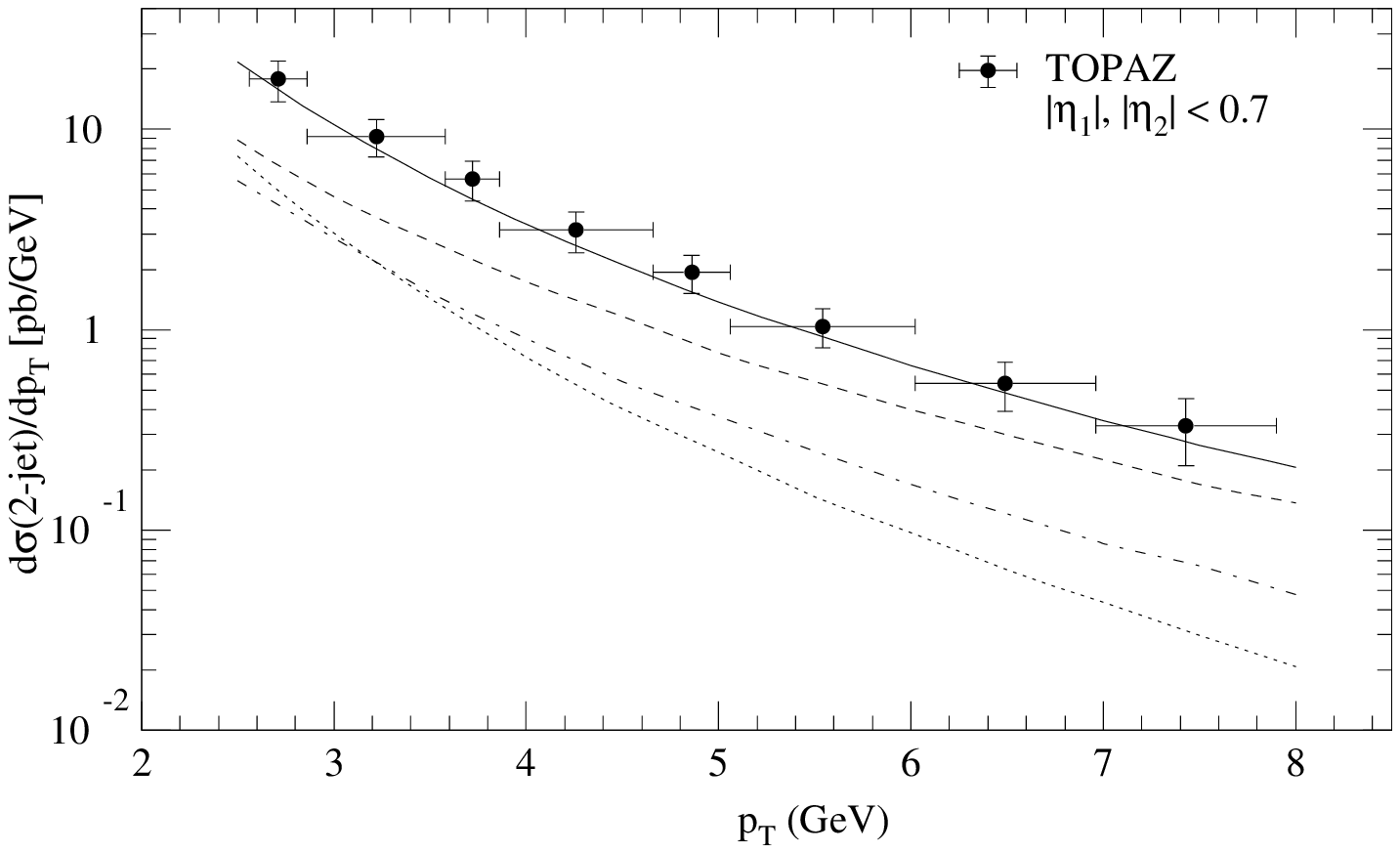, height=10cm}}
 \end{picture}
 \caption{\ }
\end{figure}
\begin{figure}[ht]
 \begin{picture}(17,10)
  \put(0,0)
   {\epsfig{file=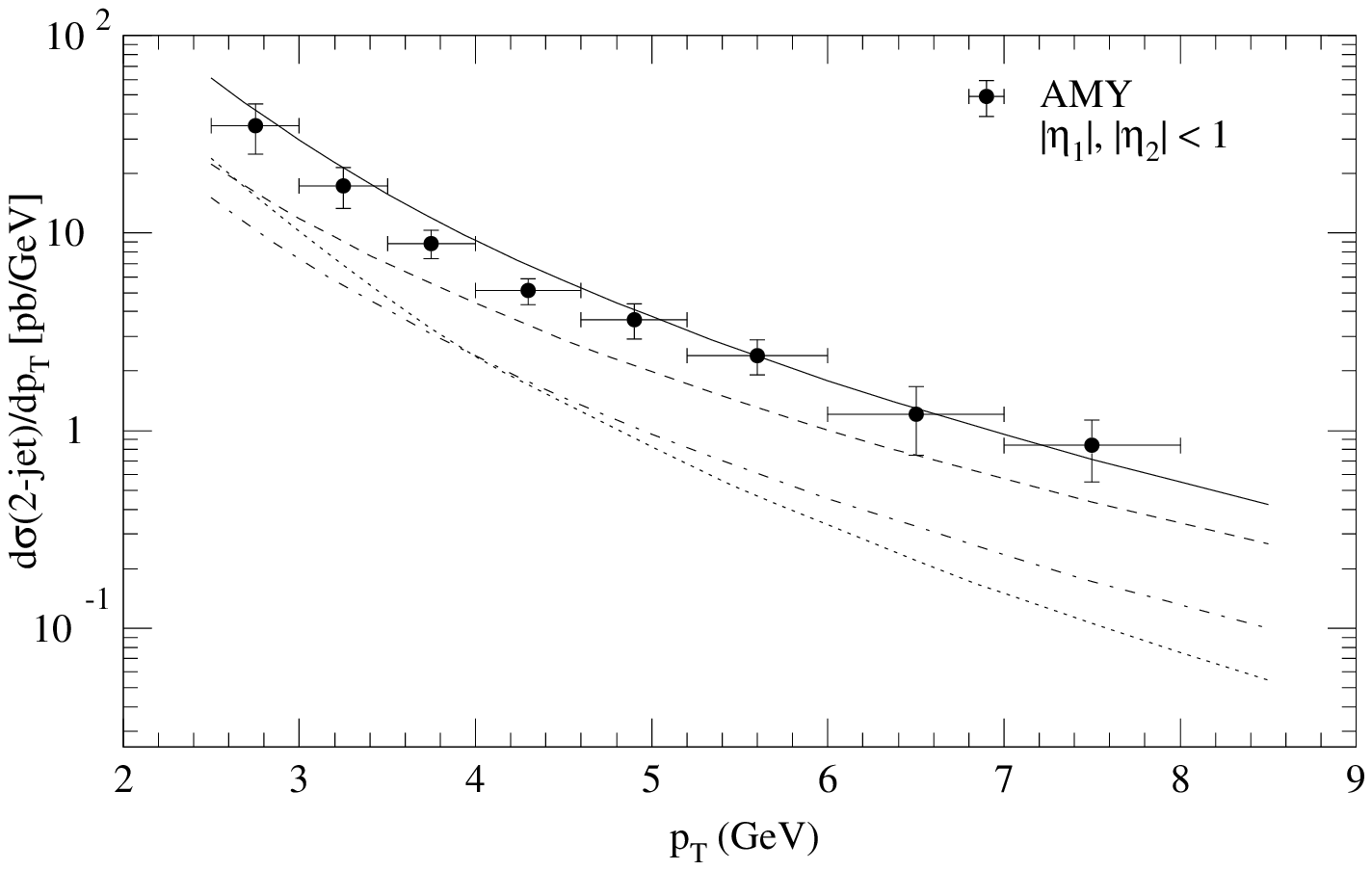, height=10cm}}
 \end{picture}
 \caption{\ }
\end{figure}
\begin{figure}[hb]
 \begin{picture}(17,10)
  \put(0,0)
   {\epsfig{file=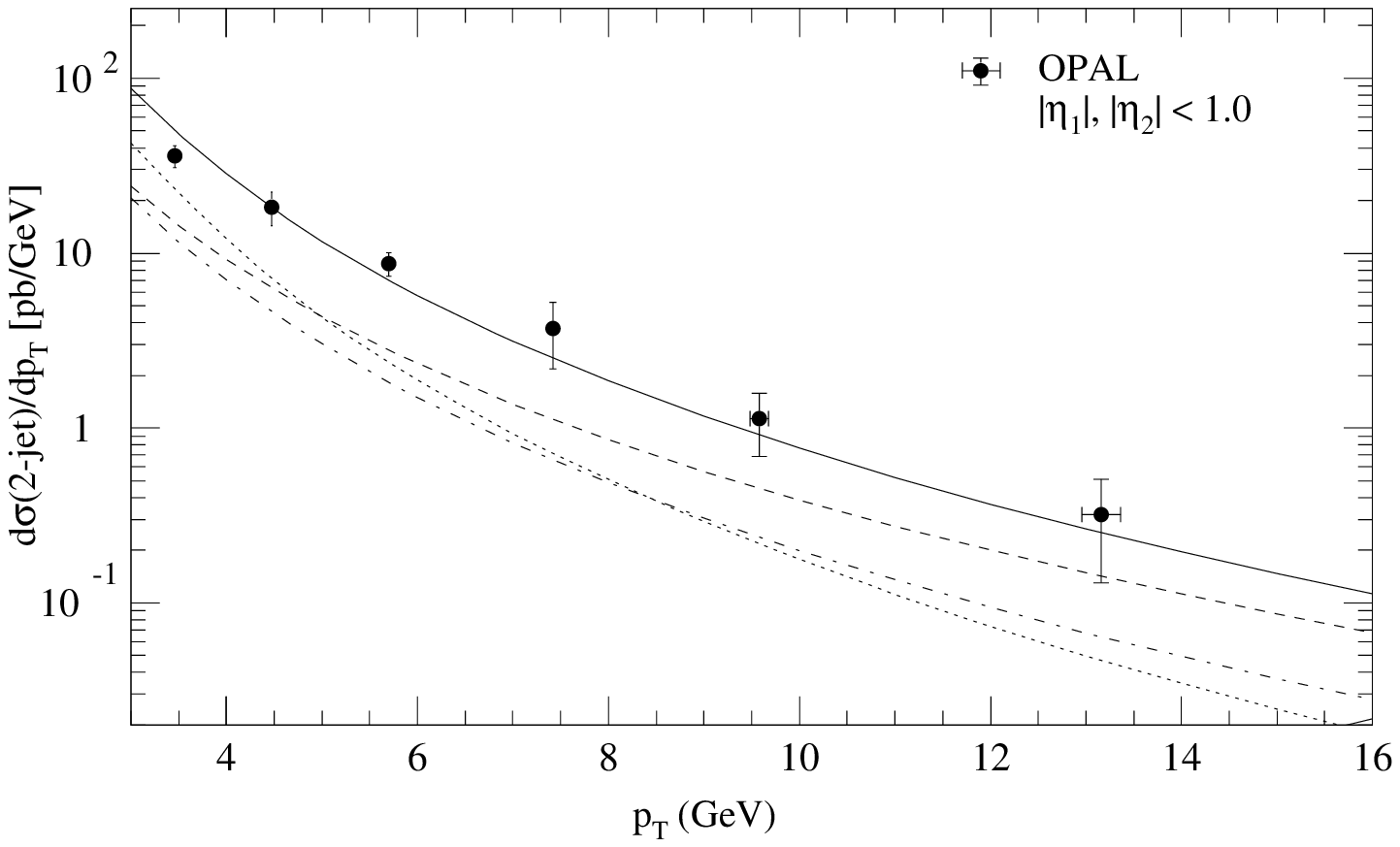, height=10cm}}
 \end{picture}
 \caption{\ }
\end{figure}
\begin{figure}[ht]
 \begin{picture}(17,22)
  \put(0,0)
   {\epsfig{file=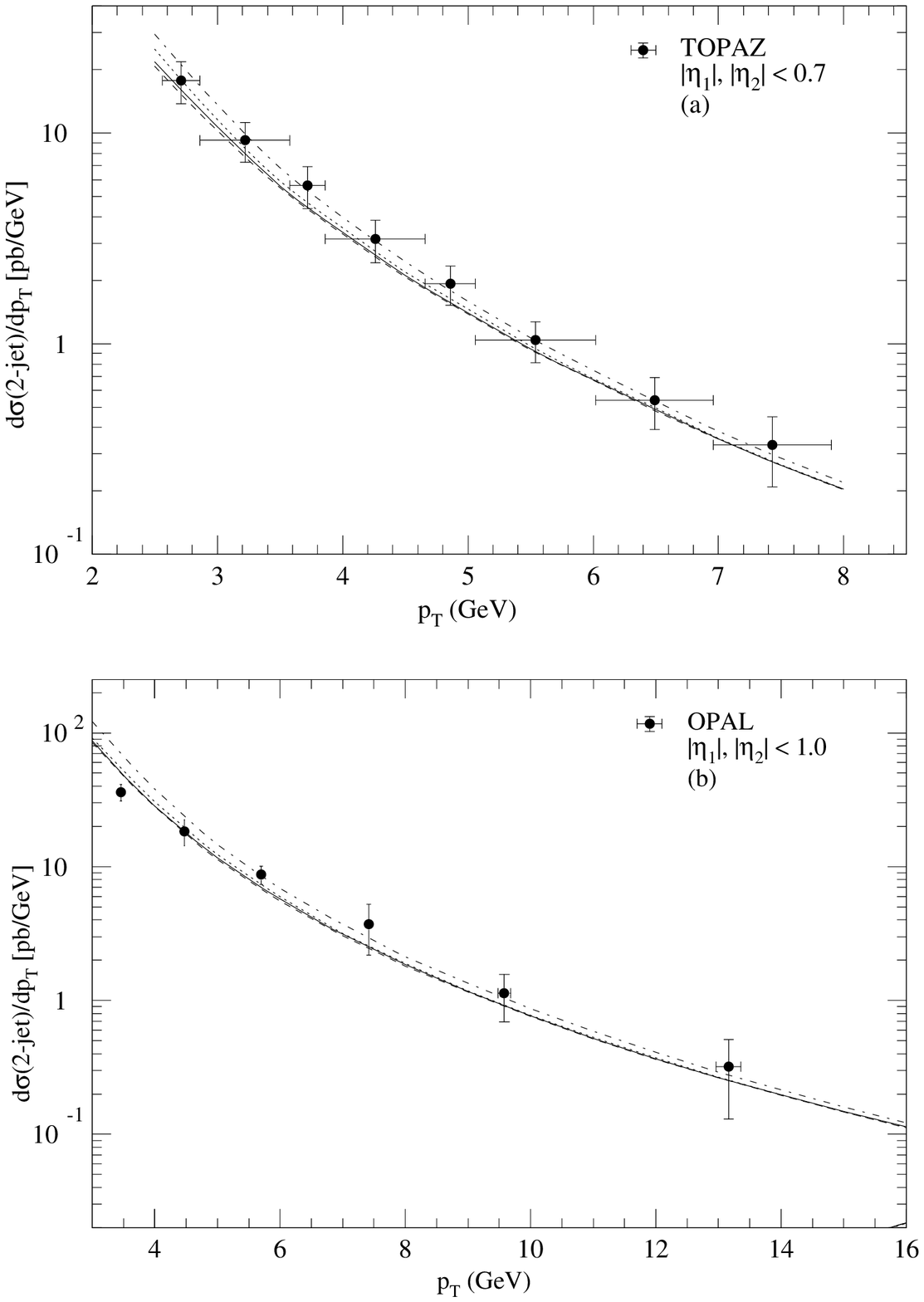, height=22cm}}
 \end{picture}
 \caption{\ }
\end{figure}
\begin{figure}[ht]
 \begin{picture}(17,22)
  \put(0,0)
   {\epsfig{file=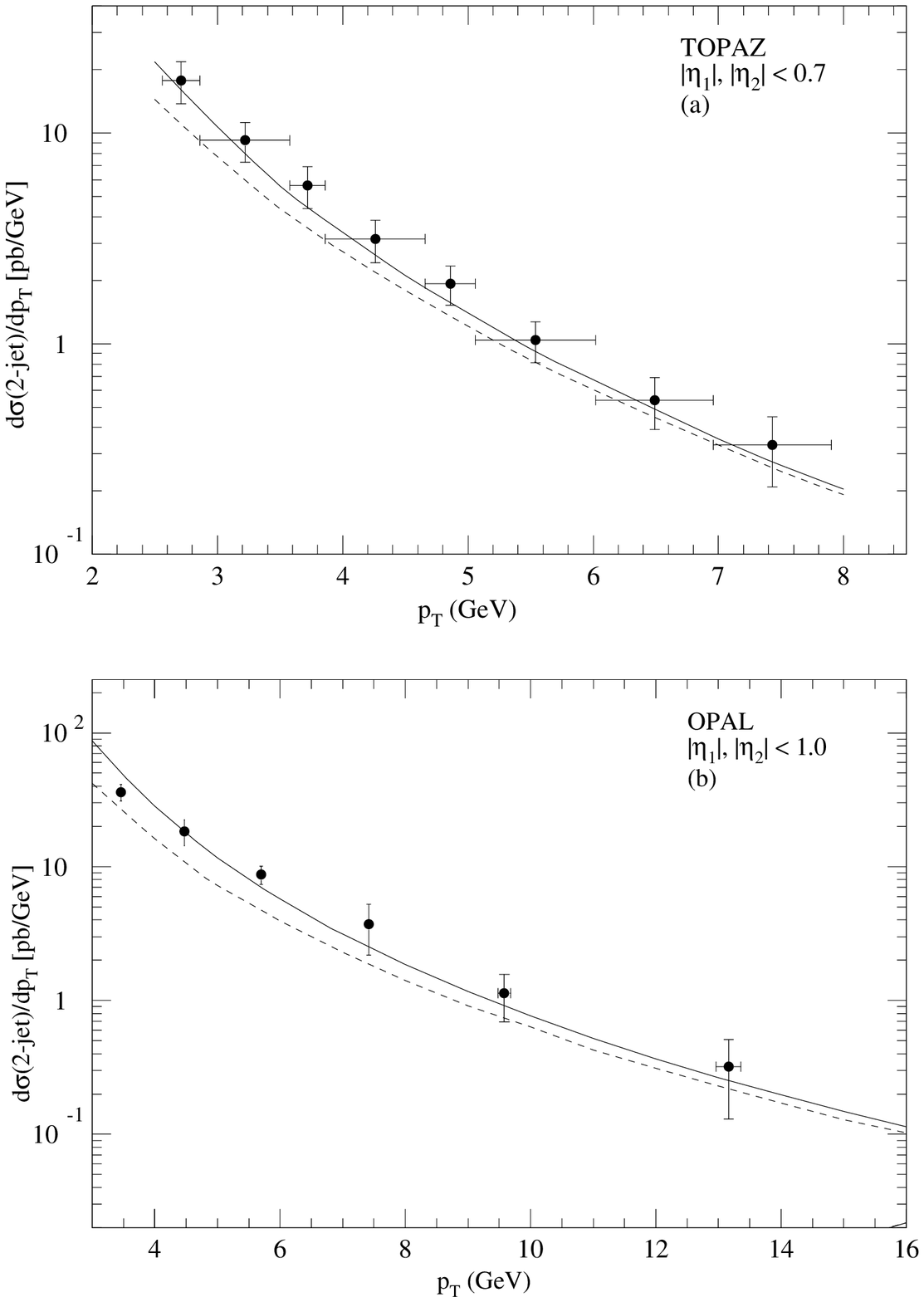, height=22cm}}
 \end{picture}
 \caption{\ }
\end{figure}
\begin{figure}[ht]
 \begin{picture}(17,22)
  \put(0,0)
   {\epsfig{file=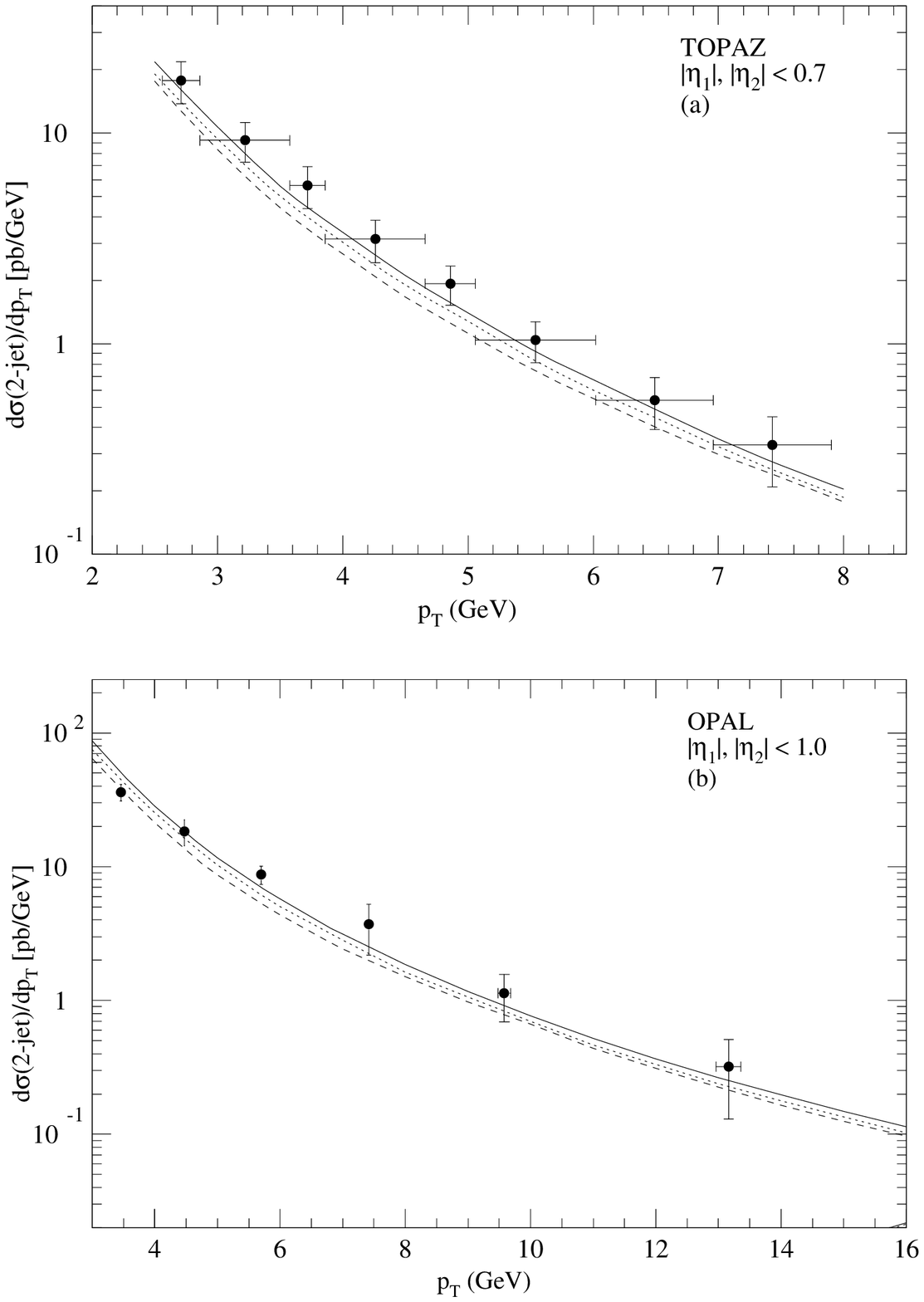, height=22cm}}
 \end{picture}
 \caption{\ }
\end{figure}

\end{document}